\documentstyle[12pt]{article}
\begin{document}
\baselineskip 24pt

\begin{center}
{\large {\bf HARMONIC GAUSS MAPS AND SELF-DUAL EQUATIONS
IN STRING THEORY}}
\end{center}

\vspace{2cm}

\begin{center}
R.Parthasarathy{\footnote{e-mail address: sarathy@imsc.ernet.in}} \\
The Institute of Mathematical Sciences \\
Madras 600 113, India \\
and \\
K.S.Viswanathan{\footnote{e-mail address: kviswana@sfu.ca}} \\
Department of Physics \\
Simon Fraser University \\
Burnaby. B.C, Canada V5A 1S6
\end{center}
\newpage
\vspace{1cm}

\noindent {\it {Abstract}}

   The string world sheet, regarded as Riemann surface, in background
$R^3$ and $R^4$ is described by the generalised Gauss map. When the Gauss map
is harmonic or equivalently for surfaces of constant mean scalar curvature, we
obtain an Abelian self-dual system, using $SO(3)$ and $SO(4)$ gauge fields
constructed in our earlier studies. This compliments our earlier result that
$h\surd g\ =\ 1$ surfaces exhibit Virasaro symmetry. The self-dual system so
obtained is compared with self-dual Chern-Simons system and a generalized
Liouville equation involving extrinsic geometry is obtained.

\vspace{0.2cm}

The world sheet in background $R^n, \ n>4$ is described by the generalized
Gauss map. It is first shown that when the Gauss map is harmonic, the scalar
mean curvature is constant. $SO(n)$ gauge fields are constructed from the
geometry of the surface and expressed in terms of the Gauss map. It is shown
that the harmonic map satisfies a non-Abelian self-dual system of equations
for the gauge group $SO(2)\times SO(n-2)$.

\newpage

\noindent {\bf I.$\ $ INTRODUCTION}

\vspace{0.5cm}

    The study of Yang-Mills connections on 2-dimensional Riemann surfaces is
of importance in string theory. The space of self-dual connections provides
a model for Teichm\"{u}ller space [1]. Clearly, the string world sheet is a
2-dimensional surface immersed in $R^n$ (For convenience we consider both the
world sheet and the target space as Euclidean). The immersion induces a metric
(the first fundamental form) on the world sheet. The second fundamental form
of the surface determines its extrinsic geometry. We [2] have developed a
formalism to study the dynamics of the world sheet conformally immersed (by
conformal immersion it is meant that the induced metric is in the conformal
gauge) in $R^3$ and $R^4$ using the generalized Gauss map [3]. The string
action can be written as a constrained Grassmannian $\sigma$-model action
and the theory is asymptotically free.

\vspace{0.5cm}

   Subsequently [4] we found a hidden Virasaro symmetry for surfaces of
constant scalar mean curvature density ($h\surd g = 1$). An action exhibiting
this symmetry has recently been constructed [5] and it is a WZNW action. The
quantum theory of this action has also been studied in [5]. It would be of
interest to know if surfaces of constant scalar mean curvature ($h$=constant)
exhibit novel properties, so that we get a better understanding of the many
facets of the string world sheet. For such surfaces, it is known that the
Gauss map from the world sheet $M$ (regarded as a Riemann surface) into the
Grassmannian $G_{2,n} \simeq SO(n)/(SO(2)\times SO(n-2))$ is harmonic for
$n$ = 3,4. It is the purpose of this paper to show that there exists an
Abelian self-dual system on such surfaces, and a non-Abelian self-dual
system for $n\ >\ 4$.

\vspace{0.5cm}

   In this paper, we first consider the Gauss map of a 2-dimensional surface
into the Grassmannian $G_{2,n}\simeq SO(n)/(SO(2)\times SO(n-2))$ for $n=3,4$.
We [4] have previously constructed $SO(n)$ connections on the string world
sheet. We project these onto $SO(2)\times SO(n-2)$ and the coset. The
projection onto $SO(2)\times SO(n-2)$ is identified as the gauge field (see
sec.III) and that on the coset as the Higg's field. Using the Euler-Lagrange
equations (harmonic map equations) for the surface, we show that this system
is self-dual. In this analysis, the harmonic map has a geometrical
interpretation. For immersion in $R^3$ and $R^4$, the Gauss map is harmonic if
the mean curvature scalar $h$ is constant [6]. We next take up immersion in
$R^n$ for $n>4$ and show explicitly that when the Gauss map is harmonic, the
mean scalar curvature of the surface is constant. The $SO(n)$ gauge fields are
constructed from the geometry of the surface. Using their projections into the
subgroup $SO(2)\times SO(n-2)$ and its compliment in $G_{2,n}$, a
non-Abelian  self-dual system is obtained when the Gauss map is harmonic.
Thus our main result is:

\vspace{0.5cm}

\noindent {\bf Theorem.1}

\vspace{0.5cm}

 {\it Let $M$ be a 2-dimensional surface conformally immersed in $R^n$. Let
$M_0$ be the Riemann surface obtained by the induced conformal structure
on $M$. Let ${\cal{G}}:M_0 \rightarrow G_{2,n}$ be the Gauss map. Let $A$
be the flat $SO(n)$ connection on $M_0$ defined by the adapted frame of
tangents and $(n-2)$ normals to $M_0$. The projections of $A$ onto
$SO(2)\times SO(n-2)$ and its orthogonal compliment in $G_{2,n}$ satisfy
self-dual system when the Gauss map $\cal{G}$ is harmonic.}

\vspace{0.5cm}

   Recently, Dunne, Jackiw, Pi and Trugenberger [7] made a systematic analysis
of the Yang-Mills non-linear Schr\"{o}dinger equation and demonstrated
self-dual Chern-Simons equations for the static configurations. Here the matter
density $\rho$ is in the adjoint representation. By choosing the Chern-Simons
gauge field in the commuting set of the Cartan subalgebra and $\Psi$ ($\rho =
-i[\Psi, {\Psi}^{\dagger}]$) in terms of the ladder operators with positive
roots, they [7] and Dunne [8] obtain Toda equations. We obtain similar results
in the case of string world sheet in background $R^n$, restricting to
world sheets of constant mean scalar curvature.

\vspace{0.5cm}

   In this way, we find that the (extrinsic) geometry of the string world sheet
described by harmonic Gauss maps
is closely related to the self-dual system of Hitchin [1] and to the static
configuration of (2+1) self-dual non-Abelian Chern-Simons theory [7,8].

\vspace{0.5cm}

\noindent {\bf II.$\ $ PRELIMINARIES}

\vspace{0.5cm}

   Consider a 2-dimensional (Euclidean) string world sheet regarded as a
Riemann surface conformally immeresed in $R^n$. The induced metric is
$g_{\alpha \beta}\ =\ {\partial}_{\alpha}X^{\mu}.{\partial}_{\beta}X^{\mu}$,
with $X^{\mu}({\xi}_1,{\xi}_2)$ as immersion coordinates ($\mu =$ 1,2,....n)
and ${\xi}_1,{\xi}_2$ as local isothermal coordinates on the surface. The
Gauss-Codazzi equations introduce the second fundamental form
$H^i_{\alpha\beta}$, $i=$ 1,2,...(n-2). Locally on the surface, we have two
tangents and (n-2) normals. The Gauss map is
\begin{eqnarray}
{\cal{G}} : M\rightarrow G_{2,n}&\simeq & SO(n)/(SO(2)\times SO(n-2)).
\end{eqnarray}
$G_{2,n}$ can be realized as a quadric $Q_{n-2}$ in $CP^{n-1}$ defined by
$\sum_{i=1}^n Z^2_i\ =\ 0$, where $Z_i$ are the homogeneous coordinates on
$CP^{n-1}$ [3]. A local tangent plane to $M_0$ is an element of $G_{2,n}$, or
equivalently a point in $Q_{n-2}$. Then [3] we have
\begin{eqnarray}
{\partial}_z X^{\mu} &=& \psi {\Phi}^{\mu},
\end{eqnarray}
where $z={\xi}_1 + i{\xi}_2$, $\bar{z} = {\xi}_1 - i{\xi}_2$, with
${\xi}_{1,2}$ as isothermal coordinates on $M_0$,  ${\Phi}^{\mu}\in
Q_{n-2}$, ${\Phi }^{\mu} {\Phi }^{\mu}=0$ and $\psi $ is a complex function
determined in terms of the geometrical properties of the surface. As not
every element of $G_{2,n}$ is a tangent plane to $M_0$, the Gauss map (2)
has to satisfy (n-2) conditions of integrability [3]. These were explicitly
derived in Ref.3 for immersion in $R^3$ and $R^4$ and by us [11] for
$R^n\  (n>4)$. We first consider immersion in $R^3$ and $R^4$.

\vspace{0.5cm}

   For immersion in $R^3$, ${\Phi}^{\mu}$ is parametrized as
\begin{eqnarray}
{\Phi}^{\mu} &=& \left[ 1-f^2, i(1+f^2), 2f \right],
\end{eqnarray}
where $f\ \in \ CP^1$. The integrability condition $f$ is,
\begin{eqnarray}
Im \left[ \frac{f_{z\bar{z}}}{f_{\bar{z}}} - \frac{2\bar{f}f_z}{1+{\mid
f\mid}^2} \right]_{\bar{z}} &=& 0.
\end{eqnarray}
The scalar mean curvature  $h (=N^{\mu} H^{\mu \alpha}_{\alpha})$ is given by
\begin{eqnarray}
(\ell n h)_z &=& \frac{f_{z\bar{z}}}{f_{\bar{z}}} - \frac{2\bar{f}f_z}{1+
{\mid f \mid}^2},
\end{eqnarray}
which is known as the Kenmotsu equation [10]. The normal $N^{\mu}$ to the
surface can be expressed in terms of $f$ as
\begin{eqnarray}
N^{\mu} &=& \frac{1}{1+{\mid f\mid}^2}\left[ f+\bar{f}, -i(f-\bar{f}),{\mid f
\mid}^2 - 1\right].
\end{eqnarray}
The energy integral of the surface is
\begin{eqnarray}
S&=&\int \frac{{\mid f_{\bar{z}}\mid }^2 + {\mid f_z \mid}^2}{(1+{\mid f
\mid}^2)^2} dz d\bar{z},
\end{eqnarray}
which is also the extrinsic curvature action $\int \surd g {\mid H\mid}^2$. The
Euler-Lagrange equation from (7) is
\begin{eqnarray}
L(f) &\equiv & f_{z\bar{z}} - \frac{2\bar{f}f_zf_{\bar{z}}}
{1+{\mid f\mid}^2} = 0.
\end{eqnarray}

\vspace{0.5cm}

   {\it {The Gauss map is said to be harmonic if $f$ satisfies the Euler-
Lagrange equations (8) and it then follows from (5) that $h$ is constant [7]}}.

\vspace{0.5cm}

   For immersion in $R^4$, we have $G_{2,4} \simeq SO(4)/(SO(2)\times
SO(2))\simeq CP^1 \times CP^1$ and so ${\Phi}^{\mu}$ is parametrized in terms
of the two $CP^1$ fields, $f_1$ and $f_2$ as
\begin{eqnarray}
{\Phi}^{\mu} &=& \left[ 1+f_1f_2, i(1-f_1f_2), f_1 - f_2, -i(f_1 + f_2)\right].
\end{eqnarray}
The Gauss map integrability conditions are
\begin{eqnarray}
Im\left[ \sum^{2}_{i=1} \frac{f_{iz\bar{z}}}{f_{i\bar{z}}} - \frac{2{\bar{f}}_i
f_{iz}}{1+{\mid f_i \mid}^2}\right]_{\bar{z}} &=& 0, \nonumber \\
\mid F_1 \mid &=& \mid F_2 \mid,
\end{eqnarray}
where $F_i = \frac{f_{i\bar{z}}}{1+{\mid f_i \mid}^2}$. There are two normals
$N^{\mu}_1\ ,\ N^{\mu}_2$ to the surface which can be written in terms of
$f_1$ and $f_2$ as
\begin{eqnarray}
N^{\mu}_1 &=& \frac{1}{2D} (A^{\mu} + {\bar{A}}^{\mu}), \nonumber \\
N^{\mu}_2 &=& \frac{1}{2iD} (A^{\mu} - {\bar{A}}^{\mu}),
\end{eqnarray}
where,
\begin{eqnarray}
D &=& \left( (1+{\mid f_1 \mid}^2)(1+{\mid f_2 \mid}^2)\right)^{\frac{1}{2}},
\nonumber \\
A^{\mu} &=& \left[f_2-{\bar{f}}_1, -i(f_2+{\bar{f}}_1), 1+{\bar{f}}_1f_2,
-i(1-{\bar{f}}_1f_2)\right], \nonumber
\end{eqnarray}
The projections of $H^{\mu \alpha}_{\alpha}$ along $N^{\mu}_1$ and
$N^{\mu}_2$ are given by [2]
\begin{eqnarray}
h_1 &=& \frac{F_1 - F_2}{2\bar{\psi}D} \nonumber \\
h_2 &=& \frac{i(F_1 + F_2)}{2\bar{\psi}D},
\end{eqnarray}
and the scalar mean curvature  $h^2\ =\ (h^2_{1} + h^2_{2})$  satisfies
the equation [2]
\begin{eqnarray}
2(\ell n h)_z &=& \sum^{2}_{i=1} \left[ \frac{f_{iz\bar{z}}}{f_{i\bar{z}}} -
\frac{2\bar{f}_if_{iz}}{1+{\mid f_i \mid}^2}\right].
\end{eqnarray}
The energy integral of the surface is
\begin{eqnarray}
S &=& \int \sum^{2}_{i=1} {\mid F_i\mid }^2 + {\mid \hat{F}_i \mid}^2,
\end{eqnarray}
where $\hat{F}_i = \frac{f_{iz}}{1+{\mid f_i \mid}^2}$. The Euler-Lagrange
equations for (14) are
\begin{eqnarray}
L(f_1) = 0 &;& L(f_2) = 0,
\end{eqnarray}
where $L(f)$ is defined in (8).
\vspace{0.5cm}

{\it The Gauss map is harmonic if $f_1$ and $f_2$ satisfy (15) and from (13)
it follows that the immersed surface has constant $h$ for immersion in
$R^4$.}

\vspace{0.5cm}

It is to be noted that $f_1$ and $f_2$ should also satisfy the second
requirement in (10) to describe the Gauss map. In Ref.4, we have considered
tangents to $M$ as ${\hat{e}}_1\ =\ \frac{1}{
\surd 2 \mid \Phi \mid}({\Phi}^{\mu} + {\bar{\Phi}}^{\mu})$ and ${\hat{e}_2}\
 =\ \frac{1}{\surd 2 i\mid \Phi \mid}({\Phi}^{\mu} - {\bar{\Phi}}^{\mu})$ along
with the (n-2) normals. Then, the local orthonormal frame $\left( {\hat{e}}_1,
{\hat{e}}_2, N^{\mu}_i \right)$ satisfies
\begin{eqnarray}
{\partial}_z {\hat{e}}_i &=& (A_z)_{ij} {\hat{e}}_j;\ \ \ \ i,j\ =\ 1\  to\  n,
\end{eqnarray}
where ${\hat{e}}_i\ =\ N^{\mu}_i$, for $i\ =\ 3$ to $n$. A similar equation
for the $\bar{z}$
derivative defines $(A_{\bar{z}})_{ij}$. $A_z$ and $A_{\bar{z}}$
transform as $SO(n,C)$ gauge fields under local $SO(n)$ transformations
of $({\hat{e}}_1, {\hat{e}}_2, N^{\mu}_i)$ which follows from (16) [4]. These
non-Abelian gauge fields are constructed from the geometrical properties of the
surface alone and so they are characteristics of the world sheet.

\vspace{0.5cm}

Using (3) and (6), it is easily verified that $A_z$ for immersion in $R^3$
is given by
\begin{eqnarray}
A_z&=&\frac{1}{1+{\mid f\mid}^2}\left[ \begin{array}{lcr}
0 & -i(f{\bar{f}}_z - \bar{f}f_z) & -(f_z + {\bar{f}}_z) \\
 & & \\
i(f{\bar{f}}_z - \bar{f}f_z) & 0 & i(f_z - {\bar{f}}_z) \\
 & & \\
f_z + {\bar{f}}_z & -i(f_z - {\bar{f}}_z) & 0
\end{array} \right]
\end{eqnarray}
Similarly, using (9),(11),(16) and, denoting $d_i=1+{\mid f_i\mid}^2$,
$m_i=f_i{\bar{f}}_{iz}-{\bar{f}}_if_{iz}$, $p_i=f_{iz} +{\bar{f}}_{iz}$, and
$q_i=f_{iz}-{\bar{f}}_{iz}$, $A_z$ for immersion in $R^4$ is obtained as
\begin{eqnarray}
A_z&=&\frac{1}{2}\left[ \begin{array}{lccr}
0 &-i(\frac{m_1}{d_1}+\frac{m_2}{d_2})&\frac{p_1}{d_1}-\frac{p_2}{d_2}&i(\frac
{q_1}{d_1} +\frac{q_2}{d_2} \\
 & & & \\
i(\frac{m_1}{d_1}+\frac{m_2}{d_2})&0&-i(\frac{q_1}{d_1}-\frac{q_2}{d_2})&\frac
{p_1}{d_1}+\frac{p_2}{d_2} \\
 & & & \\
-(\frac{p_1}{d_1}-\frac{p_2}{d_2})&i(\frac{q_1}{d_1}-\frac{q_2}{d_2})&0&i(\frac
{m_1}{d_1}-\frac{m_2}{d_2}) \\
 & & & \\
-i(\frac{q_1}{d_1}+\frac{q_2}{d_2})&-(\frac{p_1}{d_1}+\frac{p_2}{d_2})&-i(\frac
{m_1}{d_1}-\frac{m_2}{d_2})&0
\end{array} \right]
\end{eqnarray}
The gauge field $A_{\bar{z}}$ can be obtained by replacing $z$ derivatives
by $\bar{z}$ derivatives and it is seen that $(A_z)^{\dagger}\ =\ -A_{\bar{z}}$
. Further, from (16) it is easily verified that the gauge fields satisfy
\begin{eqnarray}
{\partial}_{\bar{z}}A_z - {\partial}_zA_{\bar{z}} + [A_z,A_{\bar{z}}]&=&0.
\end{eqnarray}

\vspace{0.5cm}

\noindent {\bf III.$\ $ HARMONIC GAUSS MAP AND SELF-DUAL SYSTEM}

\vspace{0.5cm}

   We now project the gauge fields constructed in the previous section onto
$SO(2)\times SO(n-2)$ and its orthogonal compliment in $G_{2,n}$ for $n \ =$
3 and 4. The general procedure is briefly outlined here. (For details see
Ref.12). Consider a $G/H$ sigma model on a two dimensional Riemann surface.
Denote the generators of the Lie algebra $L_G$ of $G$ by $L(\tilde{\sigma})$,
$\tilde{\sigma}=$ 1,2,...[G] and those of $L_H$ of $H$ by $L(\bar{\sigma})$;
$\bar{\sigma}$= 1,2,....[H]; [H]$<$[G]. The remaining generators of $L_G$ will
be denoted by $L(\sigma)$. Consider a local gauge group associated with $G$. We
have
\begin{eqnarray}
M\ \ni \ (z,\bar{z})&\stackrel{g}{\rightarrow}& g(z,\bar{z}) \in G,
\end{eqnarray}
and introduce,
\begin{eqnarray}
{\omega}_{\alpha}(g) &=& g^{\dagger}{\partial}_{\alpha}g.
\end{eqnarray}
The field strength associated with ${\omega}_{\alpha}(g)$ is zero. In fact,
${\omega}_{\alpha}(g)$ is same as $-A_z$ and $-A_{\bar{z}}$ and  is equivalent
to (16) with $g(z,\bar{z})$ as the $n\times n$ matrix formed by the two tangent
vectors ${\hat{e}}_1$ and ${\hat{e}}_2$ and the $(n-2)$ normals $N^{\mu}_i$.
Under a local gauge transformation generated by $u(z,\bar{z})\in H$, we have
\begin{eqnarray}
g(z,\bar{z})&\rightarrow & g(z,\bar{z})u(z,\bar{z}), \nonumber \\
{\omega}_{\alpha}(g)\rightarrow
{\omega}_{\alpha}(gu)&=&u^{\dagger}{\omega}_{\alpha}(g)u +
u^{\dagger}{\partial}_ {\alpha}u.
\end{eqnarray}
Thus $-A_z$ and $-A_{\bar{z}}$ transform as gauge fields under $SO(2)\times
SO(n-2)$ gauge transformation. The projection of ${\omega}_{\alpha}(g)$ onto
$L_H$ and its orthogonal compliment are
\begin{eqnarray}
a_{\alpha}(g)&=&L(\bar{\sigma}) tr (L(\bar{\sigma}){\omega}_{\alpha}(g)),
\nonumber \\
b_{\alpha}(g)&=&L(\sigma) tr (L(\sigma){\omega}_{\alpha}(g)),
\end{eqnarray}
and it is straightforward to verify that under (22),
\begin{eqnarray}
a_{\alpha}(g)\rightarrow a_{\alpha}(gu) &=& u^{\dagger}a_{\alpha}u +
u^{\dagger}{\partial}_{\alpha}u, \nonumber \\
b_{\alpha}(g)\rightarrow b_{\alpha}(gu) &=& u^{\dagger}{\partial}_{\alpha}u.
\end{eqnarray}
{\it So, $a_{\alpha}(g)$ transforms as a gauge field under local gauge
transformations  belonging to $H$ and $b_{\alpha}(g)$ transforms
homogeneously.}

\vspace{0.5cm}

   Now we consider immersion in $R^3$. The $SO(3)$ gauge fields $A_z$ and
$A_{\bar{z}}$ in (17) are projected onto $SO(2)$ and its orthogonal compliment
in $G_{2,3}$. Denoting the anti-Hermitian generators of $SO(3)$ as
$T_1,T_2,T_3$; $[T_1,T_2]\ =\ T_3$, (cyclic), we have
\begin{eqnarray}
a_z &=& \frac{1}{2} T_3 tr (T_3 A_z), \nonumber \\
b_z &=& \frac{1}{2} T_1 tr (T_1 A_z) + \frac{1}{2} T_2 tr (T_2 A_z).
\end{eqnarray}
The gauge group $H$ in (22) is $SO(2)\sim U(1)$. Similar projections for
$A_{\bar{z}}$ are made. It can be verified that $a_z\ +\ b_z\ =\ -A_z\ =\
g^{\dagger}{\partial}_z g$. Here we have a flat connection which is
decomposed as $a_z$ and $b_z$ (similarly for $A_{\bar{z}}$). Then (19)
leads to
\begin{eqnarray}
{\partial}_z a_{\bar{z}} - {\partial}_{\bar{z}} a_z + [a_z, a_{\bar{z}}] +
[b_z, b_{\bar{z}}] &=& 0, \nonumber \\
{\partial}_{\bar{z}} b_z + [a_{\bar{z}}, b_z] &=& {\partial}_z b_{\bar{z}} +
[a_z, b_{\bar{z}}],
\end{eqnarray}
where we have made use of the group structure underlying (25), viz, the
first equation in (26) is in the Cartan subalgebra while the second in $T_1$
and $T_2$ directions: hence both must separately vanish. The second equation in
(26) gives the self-dual property {\it if each side vanishes}. This we shall
prove by the equations of motion (15), namely for harmonic Gauss map.
Explicitly, from (17), we find
\begin{eqnarray}
a_z &=& \frac{1}{1+{\mid f\mid }^2}\left[ \begin{array}{lcr}
0 & i(f{\bar{f}}_z-\bar{f}f_z) & 0 \\
 & & \\
-i(f{\bar{f}}_z-\bar{f}f_z) & 0 & 0 \\
 & & \\
0 & 0 & 0
\end{array} \right],
\end{eqnarray}
and
\begin{eqnarray}
b_z &=& \frac{1}{1+{\mid f\mid}^2}\left[ \begin{array}{lcr}
0 & 0 & f_z + {\bar{f}}_z \\
 & & \\
0 & 0 & -i(f_z - {\bar{f}}_z) \\
 & & \\
-(f_z+{\bar{f}}_z) & i(f_z - {\bar{f}}_z) & 0
\end{array} \right].
\end{eqnarray}
$a_{\bar{z}}\ =\ - {a_z}^{\dagger}$ ; $b_{\bar{z}}\ =\ -{b_z}^{\dagger}$.
Then we find when $L(f) \ =\ 0$
\begin{eqnarray}
{\partial}_{\bar{z}}b_z + [a_{\bar{z}}, b_z] &=& 0,
\end{eqnarray}
Thus for {\it harmonic Gauss maps}, we have the following self-dual system
\begin{eqnarray}
{\partial}_z a_{\bar{z}} - {\partial}_{\bar{z}}a_z + [a_z, a_{\bar{z}}] + [b_z,
b_{\bar{z}}] &=& 0, \nonumber \\
{\partial}_{\bar{z}}b_z + [a_{\bar{z}}, b_z] &=& 0, \nonumber \\
{\partial}_zb_{\bar{z}} + [a_z, b_{\bar{z}}] &=& 0.
\end{eqnarray}
Note that $a_z$ transforms as an $SO(2)$ gauge field while $b_z$ transform
homogeneously and so, $b_z$ is identified with the Higg's field in Ref.1.
The self-dual system (30) is also equivalent to the static self-dual
Chern-Simons system [7,8] if we identify the matter density $\rho $ as
$-i[b_z, {b_z}^{\dagger}]$ which lies in the Cartan subalgebra of $SO(3)$.

\vspace{0.5cm}

   Next consider surfaces in $R^4$. Here we have the Grassmannian,
$G_{2,4}\simeq SO(4)/(SO(2)\times SO(2))$. We choose $T_1$ to $T_6$ as
generators of $SO(4)$ such that $[T_1, T_2]\ =\ T_3$, cyclic; $[T_4,T_5]\ =\
T_6$, cyclic; and $[T_i,T_j]\ =\ 0$ for $i\ =\ 1,2,3$; $j\ =\ 4,5,6$. The
explicit form of $A_z$ has been given in (18) and the projection of $A_z$
and $A_{\bar{z}}$ onto $SO(2)\times SO(2)$ and its compliment in $G_{2,4}$ are
\begin{eqnarray}
a_z&=& T_3 tr (T_3 A_z) + T_6 tr (T_6 A_z), \nonumber \\
b_z &=& T_1 tr (T_1A_z) + T_2 tr (T_2 A_z) + T_4 tr (T_4 A_z) + T_5 tr (T_5
A_z).
\end{eqnarray}
Equations similar to (26) readily follow from (18). The explicit forms of $a_z$
and $b_z$ are not displayed as the procedure is straightforward. The self-dual
property can be verified for harmonic maps by computing each side of the second
equation in (26) for immersion in $R^4$. Introducing
\begin{eqnarray}
{\cal L}(f_i) &=& \frac{(L(f_i) + \bar{L}(f_i))}{1+{\mid f_i\mid}^2}, \nonumber
\\
{\cal L}'(f_i)&=& \frac{(L(f_i) - \bar{L}(f_i))}{1+{\mid f_i\mid}^2}, \nonumber
\\
{\cal S} &=& {\cal L}(f_1) + {\cal L}(f_2), \nonumber \\
{\cal D} &=& {\cal L}(f_1) - {\cal L}(f_2), \nonumber \\
{\cal S}' &=& {\cal L}'(f_1) + {\cal L}'(f_2), \nonumber \\
{\cal D}' &=& {\cal L}'(f_1) - {\cal L}'(f_2),
\end{eqnarray}
for $i\ =\ 1,2$ and where $L(f)$ is defined in (8), we find
\begin{eqnarray}
{\partial}_{\bar{z}}b_z+[a_{\bar{z}},b_z]&=&\frac{1}{2}\times
   \left[ \begin{array}{lccr}
0 & 0 & -{\cal D} & -i{\cal S}' \\
 & & & \\
0 & 0 & i{\cal D}' &-{\cal S} \\
 & & & \\
{\cal D} & -i{\cal D}' &0 &0 \\
 & & & \\
i{\cal S}' & {\cal S}& 0 &0
\end{array} \right].
\end{eqnarray}
It can be seen that when the Gauss map is harmonic (the Euler-Lagrange
equations of motion (15) are satisfied) it follows that
${\partial}_{\bar{z}}b_z\ +\ [a_{\bar{z}}, b_z]=0$, which is the self-dual
equation. It is pertninent to note that $b_z$ transforms homogeneously under
the local $SO(2)\times SO(2)$ gauge transformation. The field $a_z$ which is
in the Cartan subalgebra $SO(2)\times SO(2)$ transforms as a gauge field. It
is important to reiterate that $a_z$ and $b_z$ are embedded in $SO(4)$.
Explicit solutions to the self-dual equations for the gauge group studied
here are given by (27) and (28) for $R^3$ and (18) and (31) for $R^4$, where
the complex functions $f_1$ and $f_2$ satisfy the equations, $L(f_1)=0$ and
$L(f_2)=0$. As harmonic Gauss map implies surfaces of constant scalar
mean curvature, it follows that the self-dual system is on surfaces of
constant $h$.

\vspace{0.5cm}

Fujii [14] examined the relationship between Toda systems and the
Grassmannian $\sigma$-models. We now examine the self-dual system on
surfaces in $R^3$. In anology with [7], the matter density
$\rho \ =\ {\rho}_3 T_3$ is
\begin{eqnarray}
{\rho}_3 &=& \frac{2(f_z{\bar{f}}_{\bar{z}} - f_{\bar{z}}{\bar{f}}_z)}
{(1+{\mid f \mid}^2)^2}.
\end{eqnarray}
We recall the following relations for Gauss map in $R^3$ [2]. Writing
$\surd g\ =\ exp( \phi)$, and  $\hat{F}\ =\ \frac{f_z}{1+{\mid f\mid}^2}$,
we have
\begin{eqnarray}
{\mid F\mid}^2&=& \frac{h}{2}H_{z\bar{z}} = \frac{h^2}{2}\exp(\phi), \nonumber
\\
{\mid \hat{F}\mid}^2&=& \frac{h}{2}
\frac{H_{zz}H_{\bar{z}\bar{z}}}{H_{z\bar{z}}}=\frac{1}{2}{\mid
H_{zz}\mid}^2\exp(-\phi), \nonumber \\
H_{z\bar{z}}&=&h\surd g = h\exp(\phi), \nonumber \\
{\rho}_3 &=& 2({\mid \hat{F}\mid}^2 - {\mid F\mid}^2),
\end{eqnarray}
where $H_{\alpha \beta}$ is the second fundamental form. It can be verified
by using the equation of motion (8) and the Gauss map relation, $(\ell
n\psi)_{\bar{z}}\ =\ -\frac{2\bar{f}f_{\bar{z}}}{(1+{\mid f\mid}^2)}$, that
${\partial}_z{\partial}_{\bar{z}}\ell n{\mid H_{zz}\mid}^2\ =\ 0$. The Gauss
curvature $R\ =\ -\exp(-\phi){\partial}_z{\partial}_{\bar{z}}\phi$. Writing the
Gauss curvature in terms of $H_{\alpha \beta}$, we have a modified Liouville
equation for extrinsic curvature as
\begin{eqnarray}
{\partial}_z{\partial}_{\bar{z}}\phi &=& -2h^2\exp(\phi) +
2\exp(-\phi)\exp({\phi}_E),
\end{eqnarray}
where ${\mid H_{zz}\mid}^2\ =\ \exp({\phi}_E)$ and
${\partial}_z{\partial}_{\bar{z}}{\phi}_E\ =\ 0$. When we consider $f$ to be
anti-holomorphic, (36) reduces to the Liouville equation for $h\ =\ 1$
\begin{eqnarray}
{\partial}_z{\partial}_{\bar{z}}\phi &=& -2\exp(\phi),
\end{eqnarray}
which is also the Toda equation for $SO(3)$.

\vspace{0.5cm}

For immersion in $R^4$, when we consider $f_1$ and $f_2$ both
anti-holomorphic, we obtain
\begin{eqnarray}
\rho &=& \frac{f_{1\bar{z}}{\bar{f}}_{1z}}{(1+{\mid f_1\mid}^2)^2} T_6 +
\frac{f_{2\bar{z}}{\bar{f}}_{2z}}{(1+{\mid f_2\mid}^2)^2} T_3.
\end{eqnarray}
The Cartan matrix for $SO(4)$ is $K_{\alpha\beta}\ =\
-2{\delta}_{\alpha\beta}$. Then we obtain
\begin{eqnarray}
{\partial}_z{\partial}_{\bar{z}}\ell n {\rho}_6 &=& -2{\rho}_6, \nonumber \\
{\partial}_z{\partial}_{\bar{z}}\ell n {\rho}_3 &=& -2{\rho}_3,
\end{eqnarray}
where ${\rho}_6$ and ${\rho}_3$ are the coefficients of $T_6$ and $T_3$
in (38). (39) is identified as the $SO(4)$ Toda system. Thus, harmonic
Gauss maps lead to $SO(3)$ and $SO(4)$ Toda system for immersion of the
world sheet in background $R^3$ and $R^4$ respectively.

\vspace{0.5cm}

\noindent{\bf IV.$\ $ HARMONIC GAUSS MAPS IN $R^n$ ($n\ >\ 4$)}

\vspace{0.5cm}

   We now consider immersion of 2-dimensional surfaces in $R^n,\ \  n>4$.
There are two reasons for this consideration. First of all, the gauge field
$a_z$ in the two cases considered (n=3 and 4) is an Abelian embedding in
$SO(3)$ and $SO(4)$. This is similar to the choice in Ref.8 and 9. We would
like to realize a non-Abelian self-dual system, which occurs when $n>4$.
Secondly, the result that harmonic Gauss map implies constant scalar mean
curvature, has been proved for immersion in $R^3$ by Ruh and Vilms [7] and
can be proved from our [2] results for $h$ and the Euler-Lagrange equations
for immersion in $R^4$. For immersion in $R^n,\ n>4$, such a result has not
yet been explicitly obtained to the best of our knowledge. In this paper we
prove this and use it to obtain the self-duality equations for harmonic
Gauss maps in $R^n,\ \ n>4$.

\vspace{0.5cm}

We recall the essential details of the Gauss map of surfaces in $R^n$ from
our earlier paper [10] and from Hoffman and Osserman [11]. ${\Phi}^{\mu}$
in $Q_{n-2}$ in (2) is parmetrized in the following manner. Let
$(z_1,z_2,\ .\ .\ .\ . z_n)$ be the homogeneous coordinates of $CP^{n-1}$.
The quadric $Q_{n-2}\in CP^{n-1}$ is defined by
\begin{eqnarray}
\sum_{k=1}^n {z_k}^2 &=& 0.
\end{eqnarray}
Let $H$ be the hyperplane in $CP^{n-1}$ defined by $H:(z_1 - iz_2)=0$. Then
$Q_{n-2}^* = Q_{n-2}\setminus \{H\}$ is biholomorphic to $C^{n-2}$ under the
correspondence [10,11]
\begin{eqnarray}
(z_1.....z_n)&=&\frac{z_1-iz_2}{2}\left[1-{{\zeta}_k}^2,i(1+{{\zeta}_k}^2),
2{\zeta}_1,.....,2{\zeta}_{n-2}\right],
\end{eqnarray}
where,
\begin{eqnarray}
{\zeta}_j &=& \frac{z_{j+2}}{z_1-iz_2},
\end{eqnarray}
for $j=1,2,....n-2$. (In (41) and in what follows we use the summation
convention that repeated indices are summed from $1$ to $n-2$, unless otherwise
stated.) Conversely, given any $({\zeta}_1,...{\zeta}_{n-2})\in C^{n-2}$,
\begin{eqnarray}
{\Phi}^{\mu}&=&\left[1-{{\zeta}_k}^2,i(1+{{\zeta}_k}^2),2{\zeta}_1,....
2{\zeta}_{n-2}\right],
\end{eqnarray}
satisfies (40) and hence defines a point in the complex quadric $Q_{n-2}$. The
Fubini-Study metric on $CP^{n-1}$ induces a metric on $Q_{n-2}$ [10] which is
computed as
\begin{eqnarray}
g_{ij}&=&\frac{4}{{\mid \Phi\mid}^2}{\delta}_{ij} + \frac{16}{{\mid
\Phi\mid}^4}
[{\zeta}_i{\bar{\zeta}}_j-{\zeta}_j{\bar{\zeta}}_i+2{\zeta}_i{\bar{\zeta}}_j
{\mid
{\zeta}_k\mid}^2-{\zeta}_i{\zeta}_j{{\bar{\zeta}}_k}^2-{\bar{\zeta}}_i{\bar
{\zeta}}_j{{\zeta}_k}^2],
\end{eqnarray}
where,
\begin{eqnarray}
{\mid \Phi\mid}^2&=&2+4{\zeta}_k{\bar{\zeta}}_k+2{{\zeta}_k}^2{{\bar{\zeta}}_m
}^2.
\end{eqnarray}
We [10] found it convenient to introduce an n-vector
\begin{eqnarray}
{A}^{\mu}_k&=& -[{\bar{\zeta}}_k+{\zeta}_k{{\bar{\zeta}}_m}^2]{\Phi}^{\mu} +
\frac{{\mid \Phi\mid}^2}{2}  {v}^{\mu}_k,
\end{eqnarray}
for $k=1,2...(n-2)$ and
\begin{eqnarray}
{v}^{\mu}_k&=&(-{\zeta}_k, i{\zeta}_k,0,0,0,..1_k,0,..),
\end{eqnarray}
where $1_k$ stands for 1 in the $(k+2)$th position. The algebraic properties of
${A}^{\mu}_k$ and ${a}^{\mu}_k$ have been established in [10]. The $(n-2)$ real
normals to the surface have been derived as
\begin{eqnarray}
{N}^{\mu}_i &=& \frac{4}{{\mid \Phi\mid}^4} (O^T)_{ij} {A}^{\mu}_j,
\end{eqnarray}
where the $(n-2)\times (n-2)$ matrix $O$ has been defined in [10].

\vspace{0.5cm}

   The $(n-2)$ complex functions ${\zeta}_i(z,\bar{z})$ where $z,\bar{z}$ are
the isothermal coordinates on $M_0$ have been shown to satisfy $(n-2)$
conditions so that they can represent the Gauss map [10]. The mean curvature
scalar $h$ of the surface has been shown to be related to Gauss map by
\begin{eqnarray}
(\ell n h)_z&=&
\frac{\sum^{n-2}_{j=1}{\zeta}_{j\bar{z}}{\zeta}_{jz\bar{z}}}{\sum^{n-2}_{j=1}
({\zeta}_{j\bar{z}})^2}-\frac{4}{{\mid
\Phi\mid}^2}\sum^{n-2}_{j=1}{\zeta}_{jz}[{\bar{\zeta}}_j+{\zeta}_j\sum^{n-2}_
{k=1}{{\bar{\zeta}}_k}^2],
\end{eqnarray}
which is the generalization of the Kenmotsu equation to immersion in $R^n$.

\vspace{0.5cm}

\noindent {\bf Theorem.2}

\vspace{0.5cm}

{\it Let $M$ be a 2-dimensional surface defined by a conformal immersion
$X:M\rightarrow R^n$. Then if the Gauss map ${\cal {G}}:M\rightarrow G_{2,n}$
is harmonic, the mean curvature scalar $h$ of $M$ is constant.}

\vspace{0.5cm}

\noindent {\it Proof}

\vspace{0.5cm}

   The Gauss map is said to be harmonic if the (n-2) complex functions
${\zeta}_i$ satisfy the Euler-Lagrange equations of the energy integral [10]
\begin{eqnarray}
{\cal {E}} &=& \int g_{ij} {\zeta}_{i\bar{z}}{\bar{\zeta}}_{jz}\ dz\wedge
d\bar{z}.
\end{eqnarray}
The above `energy integral' is also the action for the extrinsic curvature of
the surface $M$, namely, $\int \surd g {\mid H\mid}^2$. The Euler-Lagrange
equations that follow from the extremum of ${\cal{E}}$ are obtained as
\begin{eqnarray}
{\zeta}_{kz\bar{z}}&=& -\frac{4}{{\mid \Phi\mid}^2} \sum^{n-2}_{i=1}
{\zeta}_{i\bar{z}} {\zeta}_{iz} ({\bar{\zeta}}_k +
{\zeta}_k\sum^{n-2}_{m=1}{{\bar{\zeta}}_m}^2) \nonumber \\
& + & \frac{4}{{\mid \Phi\mid}^2}\sum^{n-2}_{i=1}({\bar{\zeta}}_i +
{\zeta}_i\sum^{n-2}_{m=1}{{\bar{\zeta}}_m}^2) [{\zeta}_{k\bar{z}}{\zeta}_{iz} +
{\zeta}_{kz}{\zeta}_{i\bar{z}}].
\end{eqnarray}
Upon using the above  equations in the expression for $(\ell n
h)_z$ in (49) it follows that the mean curvature scalar $h$ of the
surface $M_0$ is constant. This completes the proof.

\vspace{0.5cm}

   We now proceed to construct $SO(n)$ gauge fields on the surface $M_0$.
The defining equation for them is (16) with ${\Phi}^{\mu}$ given by (43) and
the $(n-2)$ normals by (48). The various components of $A_z$ are given below.
\begin{eqnarray}
A_z &=& \left[\begin{array}{lc|c}
0 & (A_z)_{12} & (A_z)_{1i}(i=3 \mbox { to } n) \\
 & & \\
(A_z)_{21} & 0 & (A_z)_{2i}(i=3 \mbox { to } n) \\
\hline
(A_z)_{31} & (A_z)_{32} & ........    \\
 & & \\
\vdots &\vdots &    (A_z)_{ij} \\
 & & \\
(A_z)_{n1} & (A_z)_{n2} &...
 \end{array}\right]
\end{eqnarray}
where,
\begin{eqnarray}
(A_z)_{12}&=& \frac{2}{i{\mid
\Phi\mid}^2}[({\zeta}_j+{{\bar{\zeta}}_j}{{\zeta}_i}^2){\bar{\zeta}}_{jz} -
({\bar{\zeta}}_j + {\zeta}_j{{\bar{\zeta}}_i}^2){\zeta}_{jz}] \nonumber \\
(A_z)_{1i}&=& \frac{2\surd 2}{{\mid \Phi\mid}^5} (O^T)_{ij}[{\mid
\Phi\mid}^2{\zeta}_{jz} -
4({\bar{\zeta}}_j+{\zeta}_j{{\bar{\zeta}}_m}^2)({\zeta}_k+{\bar{\zeta}}_k
{{\zeta}_q}^2){\bar{\zeta}}_{kz} \nonumber \\
&+& {\mid
\Phi\mid}^2(2{\zeta}_j{\bar{\zeta}}_k{\bar{\zeta}}_{kz}+{\bar{\zeta}}_{jz})]
\nonumber  \\
(A_z)_{2i}&=&\frac{2\surd 2}{i{\mid \Phi\mid}^5} (O^T)_{ij}[{\mid \Phi\mid}^2
{\zeta}_{jz} +
4({\bar{\zeta}}_j+{\zeta}_j{{\bar{\zeta}}_m}^2)({\zeta}_k+{\bar{\zeta}}_k
{{\zeta}_q}^2){\bar{\zeta}}_{kz} \nonumber \\
&-& {\mid \Phi\mid}^2(2{\zeta}_j{\bar{\zeta}}_k{\bar{\zeta}}_{kz} +
{\bar{\zeta}}_{jz})] \nonumber \\
(A_z)_{ij}&=&-\frac{1}{{\mid \Phi\mid}^2}{\partial}_z({\mid
\Phi\mid}^2){\delta}_{ij} + \frac{4}{{\mid
\Phi\mid}^4}(O^T)_{jk}{\partial}_zO_{ki}  \nonumber \\
&+&\frac{16}{{\mid
\Phi\mid}^6}({\bar{\zeta}}_k+{\zeta}_k{{\bar{\zeta}}_q}^2){\zeta}_{mz}[(O^T)
_{jk}(O^T)_{im}-(O^T)_{jm}(O^T)_{ik}],
\end{eqnarray}

\vspace{0.5cm}

$A_{\bar{z}}$ can be obtained by replacing the $z$-derivatives by $\bar{z}$
derivatives. They together define the $SO(n)$ gauge fields on the surface
$M_0$. We now project them on to $SO(2)\times SO(n-2)$ and its orthogonal
compliment in $G_{2,n} \simeq SO(n)/(SO(2)\times SO(n-2))$. Denoting these
projections by $a_z$ and $b_z$ respectively (and similarly for $a_{\bar{z}}$
and $b_{\bar{z}}$), we have
\begin{eqnarray}
a_z &=&- \left[ \begin{array}{lc|c}
0 & (A_z)_{12} & 0(13 \mbox { to } 1n) \\
 & & \\
(A_z)_{21} & 0 & 0(23 \mbox { to } 2n) \\
\hline
 0 & 0&   \\
  & & \\
\vdots &\vdots   &  (A_z)_{ij}   \\
 & & \\
 0 & 0&
\end{array}\right]
\end{eqnarray}
and
\begin{eqnarray}
b_z &=& -\left[ \begin{array}{lc|c}
0 & 0 & (A_z)_{1i}(i=3 \mbox { to } n) \\
 & & \\
0 & 0 & (A_z)_{2i}(i=3 \mbox { to } n) \\
\hline
(A_z)_{31}&(A_z)_{32}&   \\
 & & \\
\vdots & \vdots &  0    \\
(A_z)_{n1}&(A_z)_{n2}&
\end{array}\right].
\end{eqnarray}

\vspace{0.5cm}

   From the general considerations described in (24), it follows that $a_z$
transform as gauge fields under local $SO(2)\times SO(n-2)$ gauge
transformation while $b_z$ transforms homogeneously. It can be verified that
$a_z$ in (54) is indeed a non-Abelian gauge field when $n>4$.
The gauge connection $a_z$ contains contributions from
both the tangent space and the normal frame to $M$ reflecting $SO(2)\times
SO(n-2)$ group structure. The orthogonal compliment $b_z$ on the otherhand
receives contributions
from interaction of tangents with the normals. As $b_z$ transforms
homogeneously under the local $SO(2)\times SO(n-2)$ gauge transformations, it
can be identified with the Higg's field. Realizing that $A_z\ =\
-(a_z \ +\ b_z)$ and using (19) we immediately obtain (26) exploiting the group
structure underlying (54) and (55).
In order to prove that second equation in (26) gives the self
duality, namely the vanishing of both the sides, we make use of the
Euler-Lagrange equation (51) or harmonic Gauss map. To prove the self-duality
equation, namely, ${\partial}_{\bar{z}}b_z\ +\ [a_{\bar{z}}\ ,\ b_z]\ =\ 0$,
when the Gauss map is harmonic, we proceed as
below. We consider $(1i)$ component of the self-duality equation for $i\geq 3$
which follows from (54) and (55).
\begin{eqnarray}
({\partial}_{\bar{z}}b_z + [a_{\bar{z}},b_z])_{1i}&=&
{\partial}_{\bar{z}}(b_z)_{1i} + (a_{\bar{z}})_{12}(b_z)_{2i} -
(b_z)_{1j}(a_{\bar{z}})_{ji},  \nonumber \\
&=& -{\partial}_{\bar{z}}(A_z)_{1i} + (A_{\bar{z}})_{12}(A_z)_{2i} -
(A_z)_{1j}(A_{\bar{z}})_{ji},\ \ \ \ \ j\ \geq\ 3,
\end{eqnarray}
where the structure of (54) and (55) have been used. Using the definition that
$ (A_z)_{1i}\ =\ N^{\mu}_i({\partial}_z{\hat{e}}_1)$ and (16), we find,
\begin{eqnarray}
({\partial}_{\bar{z}}b_z +
[a_{\bar{z}},b_z])_{1i}&=&(A_{\bar{z}})_{12}(A_z)_{2i} +
(A_z)_{12}(A_{\bar{z}})_{2i} -
N^{\mu}_i({\partial}_z{\partial}_{\bar{z}}{\hat{e}}_1).
\end{eqnarray}

\vspace{0.5cm}

   The expressions for $(A_{\bar{z}})_{12}$ , $(A_z)_{2i}$ , $(A_z)_{12}$ and
$(A_{\bar{z}})_{2i}$ have been given in (53). The quantity
$N^{\mu}_i{\partial}_z {\partial}_{\bar{z}}{\hat{e}}_1$ is calculated using
(43) to be
\begin{eqnarray}
N^{\mu}_i{\partial}_z{\partial}_{\bar{z}}{\hat{e}}_1 &=& \frac{4}{{\mid
\Phi\mid}^4} (O_T)_{ik} [ {\partial}_{\bar{z}}(\frac{1}{\surd 2\mid
\Phi\mid})(A^{\mu}_k{\partial}_z{\Phi}^{\mu} +
A^{\mu}_k{\partial}_z{\bar{\Phi}}^{\mu}) \nonumber \\
&+& {\partial}_z (\frac{1}{\surd 2 \mid \Phi \mid})
(A^{\mu}_k{\partial}_{\bar{z}}{\Phi}^{\mu} +
A^{\mu}_k{partial}_{\bar{z}}{\bar{\Phi}}^{\mu}) \nonumber \\
&+& (\frac{1}{\surd 2\mid
\Phi\mid})(A^{\mu}_k{\partial}_z{\partial}_{\bar{z}}{\bar{\Phi}}^{\mu} +
A^{\mu}_k {\partial}_z{\partial}_{\bar{z}}{\Phi}^{\mu})].
\end{eqnarray}
Using the expressions for $A^{\mu}_k$ in (46) and ${\Phi}^{\mu}$ in (43), the
above quantity has been evaluated using,
\begin{eqnarray}
A^{\mu}_k{\partial}_z{\Phi}^{\mu} &=& {\mid \Phi \mid }^2{\zeta}_{kz},
\nonumber \\
A^{\mu}_k{\partial}_{\bar{z}}{\Phi}^{\mu} &=& {\mid \Phi \mid
}^2{\zeta}_{k\bar{z}}, \nonumber \\
A^{\mu}_k{\partial}_z{\bar{\Phi }}^{\mu} &=& -4({\bar{\zeta}}_k +
{\zeta}_k{{\bar{\zeta}}_q}^2)({\zeta}_j+{\bar{\zeta}}_j{{\zeta}_m}^2)
{\bar{\zeta}}_{jz} + {\mid \Phi \mid }^2({\bar{\zeta}}_{kz} +
2{\zeta}_k{\bar{\zeta}}_m{\bar{\zeta}}_{mz}) \nonumber \\
A^{\mu}_k{\partial}_{\bar{z}}{\bar{\Phi}}^{\mu} &=& -4({\bar{\zeta}}_k +
{\zeta}_k {\bar{\zeta}}^2_{q})({\zeta}_j +
{\bar{\zeta}}_j{\zeta}^2_{m}){\bar{\zeta}}_{j\bar{z}} + {\mid \Phi \mid
}^2({\bar{\zeta}}_{k\bar{z}} +
2{\zeta}_k{\bar{\zeta}}_m{\bar{\zeta}}_{m\bar{z}}) \nonumber \\
A^{\mu}_k{\partial}_z{\partial}_{\bar{z}}{\Phi}^{\mu} &=& 4({\bar{\zeta}}_k +
{\zeta}_k{\bar{\zeta}}^2_m){\zeta}_{qz}{\zeta}_{q\bar{z}} + {\mid \Phi \mid }^2
{\zeta}_{kz\bar{z}} \nonumber \\
A^{\mu}_k{\partial}_z{\partial}_{\bar{z}}{\bar{\Phi}}^{\mu} &=&
-4({\bar{zeta}}_k + {\zeta}_k{\bar{\zeta}}^2_m)[({\zeta}_j +
{\bar{\zeta}}_j{\zeta}^2_q){\bar{\zeta}}_{jz\bar{z}} + {\zeta}^2_q
{\bar{\zeta}}_{j\bar{z}}{\bar{\zeta}}_{jz}] \nonumber \\
& + & {\mid \Phi \mid }^2[2{\zeta}_k{\bar{\zeta}}_{j\bar{z}}{\bar{\zeta}}_{jz}
+ 2{\zeta}_k{\bar{\zeta}}_j{\bar{\zeta}}_{jz\bar{z}} +
{\bar{\zeta}}_{kz\bar{z}}],
\end{eqnarray}
and the Euler-Lagrange equations of motion (51) (harmonic map requirement) for
the $z\bar{z}$-derivatives of $\zeta$'s. We then find,
\begin{eqnarray}
({\partial}_{\bar{z}}b_z + [a_{\bar{z}} , b_z])_{1i} &=& 0,
\end{eqnarray}
which is the required self-duality equation. Similarly the other components
have been verified. This proves our main result that for immersed surfaces in
$R^n$ for $n>4$, the surface $M$ admits self-dual system when the
Gauss map is harmonic. Explicit solutions to the self-dual equations for the
gauge group $SO(2)\times SO(n-2)$ are given by (54) and (55), where the complex
functions $\zeta$'s satisfy the equation of motion (51).

\vspace{0.5cm}

\noindent {\bf VI.CONCLUSIONS}

\vspace{0.5cm}

   We have considered the string world sheet regarded as a Riemann surface
immersed in $R^n$. The immersion is described by the Gauss map. For $n=3$ and
4, when the Euler-lagrange equations following from the extrinsic curvature
action are satisfied, the Gauss map is harmonic and the mean curvature scalar
of the immersed surface is constant. For such a class of surfaces, we have made
use of the $SO(3)$ and $SO(4)$ two dimensional gauge fields constructed by us
[4] and projected them onto the subgroup and its orthogonal compliment in the
Grassmannian $G_{2,3}$ and $G_{2,4}$. The projection into the subgroup
transforms as a gauge field belonging to the subgroup $SO(2)$ and
$SO(2)\times SO(2)$,
while the compliment
transforms homogeneously. By identifying the compliment with the complex Higg's
field, we are able to prove the existence of solutions to Hitchin's self-dual
equation for the constant $h$ immersions in $R^3$ and $R^4$. This study
compliments our earlier result that $h\surd g$ = 1, surfaces exhibit Virasaro
symmetry.

\vspace{0.5cm}

   The self-dual system so obtained for harmonic maps is compared with the
self-dual Chern-Simons system. A generalized Liouville equation involving
extrinsic geometry is obtained. As a particular case, when the map is
anti-holomorphic the familiar Toda equations are obtained.

\vspace{0.5cm}

   We have generalized the results to conformal immersion of 2-dimensional
surfaces in $R^n$, using the results of the generalized Gauss map. We prove
that the surface has constant mean curvature when the Gauss map is harmonic.
This harmonicity condition or Euler-Lagrange equation is used to show that for
such surfaces, there exists Hitchin's self-dual system. It is to be noted
that the procedure to construct self-dual system for harmonic Gauss maps
here is a variant of the one by Donaldson [14]. In particular, the gauge
fields are constructed from the geometry of the surface itself (and so
characteristic of the surface and not external) and the self-dual gauge
fields $a_{z}$ and $a_{\bar{z}}$ belong to the gauge group $SO(2)\times
SO(n-2)$ while the fields $b_{z}$ and $b_{\bar{z}}$ belong to the
compliment in $G_{2,n}$, transform homogeneously under $SO(2)\times
SO(n-2)$ gauge transformation.

\vspace{0.5cm}

   The general action for the string theory will be a sum of the Nambu-Goto
action and the action involving extrinsic geometry. When the theory is
described in terms of the Gauss map, we have earlier [2] noted that both the
actions can be expressed as Grassmannian sigma model. Explicitly,
\begin{eqnarray}
S &=& S_{NG} + S_{Extrinsic} \nonumber \\
&=&\sigma \int \surd g dz\wedge d\bar{z} +\frac{1}{{\alpha}_0}
\int \surd g {\mid H \mid }^2
dz\wedge d\bar{z}, \nonumber \\
&=&\sigma \int \frac{1}{h^2} g_{ij}{\zeta}_{i\bar{z}}{\bar{\zeta}}_{jz}\
\ dz\wedge d\bar{z} + \frac{1}{{\alpha}_0}
\int g_{ij}{\zeta}_{i\bar{z}}{\bar{\zeta}}_{jz}\  dz\wedge
d\bar{z}.
\end{eqnarray}
In general the mean curvature scalar $h$ will be a (real) function of
(z,$\bar{z}$)
and so the study of the action $S$ will be complicated since we have a
space-dependent coupling for $S_{NG}$ in this frame work. When the Gauss
map is harmonic (surfaces of constant mean scalar curvature), it is easy to
see that the total action is just a Grassmannian sigma model with one
effective
coupling constant. Since the classical equations of motion for this action are
identical to (51), it is possible to study the quantization in the background
field method.

\vspace{0.5cm}

\noindent {\bf Acknowledgements}

\vspace{0.5cm}

We are thankful to Prof.I.Volovich and Prof.I.Y.Arefeva for useful discussions.
One of us (R.P) wishes to thank Prof.C.S.Seshadri for valuable discussions.
This work has been supported by an operating grant (K.S.V) from the Natural
Sciences and Engineering Council of Canada. R.P thanks the Department of
Physics, Simon Fraser University, for the hospitality during the summer 1993.

\vspace{0.5cm}

\noindent {\bf References}

\begin{enumerate}
\item N.J.Hitchin, Proc.London.Math.Soc.(3){\bf 55},59(1987).

\item K.S.Viswanathan,R.Parthasarathy and D.Kay,Ann.Phys.(N.Y){\bf \\
206}237(1991).

\item D.A.Hoffman and R.Osserman,J.Diff.Geom.{\bf 18},733(1983); \\
Proc.London.Math.Soc.(3){\bf 50},21(1985).

\item R.Parthasarathy and K.S.Viswanathan,Int.J.Mod.Phys.{\bf
A7},317(1992).

\item K.S.Viswanathan and R.Parthasarathy, Ann.Phys.(N.Y)(submitted).

\item E.A.Ruh and J.Vilms,Trans.Amer.Math.Soc.{\bf 149},569(1979).

\item G.V.Dunne,R.Jackiw,S-Y.Pi and C.A.Trugenberger, Phys.Rev.{\bf D43}, \\
1332(1991).

\item G.V.Dunne, Comm.Math.Phys.{\bf 150},519(1992).

\item K.Kenmotsu, Math.Ann.{\bf 245},89(1979).

\item R.Parthasarathy and K.S.Viswanathan,Int.J.Mod.Phys.{\bf \\
A7},2819(1992).

\item D.A.Hoffman and R.Osserman,Mem.Amer.Math.Soc.No.236 (1980).

\item A.P.Balachandran,A.Stern and G.Trahern,Phys.Rev.{\bf D19},2416(1979).

\item K.Fujii,Lett.Math.Phys.{\bf 25},203(1992).

\item S.K.Donaldson, Proc.London.Math.Soc.(3) {\bf 55}, 127(1987).
\end{enumerate}
\end{document}